\documentclass[acus]{JAC2003}

\usepackage{graphicx}
\usepackage{booktabs}
\usepackage{multirow}

\begin{document}
\title{MEASUREMENTS OF INCLUSIVE W/Z PRODUCTION CROSS SECTIONS AT CMS AND W/Z AS A LUMINOMETER}

\author{Jeremy Werner, Princeton University, Princeton, NJ USA \\
  on behalf of the CMS Collaboration, CERN, Geneva, Switzerland}
\bigskip
\maketitle
\vskip 1cm
\bigskip
\bigskip
\begin{abstract}
  Leptonic decays of W/Z bosons provide the first electroweak precision measurements at the Large Hadron Collider (LHC).
 The results of measurements of inclusive $\mathrm{W}$ and $\mathrm{Z}$ boson production cross sections in pp 
 collisions at $\sqrt{s}=7~\mathrm{TeV}$ are presented~\cite{theArticle}, based on $2.9~\mathrm{pb}^{-1}$ of data recorded by the Compact
Muon Solenoid (CMS) detector at the LHC. The measurements, performed in the electron and muon decay channels,
are combined to give $\sigma( \mathrm{pp} \rightarrow \mathrm{W}X ) 
\times {\cal{B}}( \mathrm{W} \rightarrow \ell \nu ) = 9.95\pm0.07\,{\textrm{(stat.)}}\pm 0.28\,{\textrm{(syst.)}}\pm1.09\,{\textrm{(lumi.)}}~\mathrm{nb}$ and $\sigma( \mathrm{pp} \rightarrow \mathrm{Z}X ) 
\times {\cal{B}}( \mathrm{Z} \rightarrow \ell^+ \ell^- ) = 0.931\pm0.026\,{\textrm{(stat.)}}\pm 0.023\,{\textrm{(syst.)}}\pm0.102\,{\textrm{(lumi.)}}~\mathrm{nb}$, where $\ell$ stands for either $\mathrm{e}$ or $\mu$.
  Theoretical predictions, calculated at the next-to-next-to-leading order (NNLO) in QCD using 
recent parton distribution functions (PDFs), are in agreement with the measured cross sections.
Hence copious production of these well understood and clean signatures suggest the use of W/Z as 
 a ``standard candle'' for measuring the luminosity at the LHC alongside the current Van der Meer (VdM) separation
 scan method.  
\end{abstract}

\section{Introduction}

Inclusive leptonic decays of W and Z bosons are benchmark physics processes at hadron colliders. These
first electroweak processes studied at the LHC allow validation of high transverse momentum
electron and muon reconstruction and identification.  In addition, precision measurements of the W/Z 
are important in testing the Standard Model more rigorously than ever before, 
constraining the PDF, and
potentially uncovering signs of new physics that could appear through radiative corrections.

 The results of the W/Z production cross section  measurements  with pp collisions at a center-of-mass energy of $7$~TeV provided by the  LHC are reported~\cite{theArticle}.
The data were collected from April through August, 2010, by the CMS experiment, and correspond to an integrated luminosity of $2.9  \pm 0.3$ pb$^{-1}$. The consistency of the results between the different leptonic decay channels and with the NNLO 
theoretical calculations suggests already 
considering the use of these electroweak boson decays as ``Standard Candles for LHC'' to calibrate 
the absolute luminosity alongside the Van deer Meer separation scan. Comparison of the systematic uncertainties between the two methods is provided.

The precision of the cross section measurements was limited by the systematic uncertainty on the luminosity (11\%).
In the very near future~\footnote{
As of the publication date of this article the luminosity uncertainty has been reduced to 4\%.}
more detailed understanding of several of
the main systematic biases will substantially reduce the uncertainty of these measurements.
The statistical uncertainty will also be reduced by about a factor of 3 once the measurement is performed on
the entire 2010 data set, corresponding to 36 pb$^{-1}$.
Conservative systematic uncertainty projections for the measurements using the full 2010 data set are provided.~\footnote{
Since the LHC Lumi Days workshop, this analysis on 36 pb$^{-1}$ has been completed~\cite{EWK10005}.}


\section{Cross Section Results for 2.9 \lowercase{pb$^{-1}$}}


Results for electron and muon decay
channels are reported separately, and then combined assuming lepton universality in W and Z decays.
The electron and muon channels are combined 
by maximizing a likelihood that accounts for the individual 
statistical and systematic uncertainties and their correlations.
For cross section measurements, 
correlations are only numerically relevant for theoretical
uncertainties, including the PDF uncertainties on the acceptance values.
For cross section ratio
measurements, the correlations of lepton efficiencies are taken into account in each lepton
channel, with other experimental uncertainties assumed uncorrelated; in the combination
of lepton channels, fully-correlated uncertainty for the acceptance factor are assumed, with
other uncertainties assumed uncorrelated.

Table~\ref{tab:xsecResults} summarizes the measured electroweak boson production cross sections, and compares them
to their theoretical NNLO predictions~\cite{NNLO1,NNLO2}.
The reported Z boson production cross
sections pertain to the invariant mass range $M_{\ell\ell} \in (60,120)$ GeV, and are corrected for
the fiducial and kinematic acceptance but not for $\gamma^{*}$ exchange.
Each cross section result in the table carries an additional uncertainty of 11\%
from the luminosity that is not listed.  

\begin{table}[hbt]
   \centering
   \caption{Summary of the production cross section times branching ratio measurements and their
theoretical predictions}
   \footnotesize
   \begin{tabular}{cccc}
       \toprule
       \multicolumn{2}{c}{\textbf{Channel}} & \textbf{$\sigma\times\mathcal{B}$ (nb) } & \textbf{NNLO (nb)} \\
       \midrule

           \multirow{3}*{W} & $e\nu$    & $10.04 \pm 0.10 (\mathrm{stat}) \pm 0.52 (\mathrm{syst})$ & \multirow{3}{*}{$10.44 \pm 0.52$} \\
                            & $\mu\nu$  & $9.92 \pm 0.09 (\mathrm{stat}) \pm 0.31 (\mathrm{syst})$ & \\
                            & $\ell\nu$ & $9.95 \pm 0.07 (\mathrm{stat}) \pm 0.28 (\mathrm{syst})$ & \\\hline
     \multirow{3}*{W$^{+}$} & $e^{+}\nu$    & $5.93 \pm 0.07 (\mathrm{stat}) \pm 0.36 (\mathrm{syst})$ & \multirow{3}{*}{$6.15 \pm 0.29$} \\
                            & $\mu^{+}\nu$  & $5.84 \pm 0.07 (\mathrm{stat}) \pm 0.18 (\mathrm{syst})$ & \\
                            & $\ell^{+}\nu$ & $5.86 \pm 0.06 (\mathrm{stat}) \pm 0.17 (\mathrm{syst})$ & \\\hline
     \multirow{3}*{W$^{-}$} & $e^{-}\nu$    & $4.14 \pm 0.06 (\mathrm{stat}) \pm 0.25 (\mathrm{syst})$ & \multirow{3}{*}{$4.29 \pm 0.23$} \\
                            & $\mu^{-}\nu$  & $4.08 \pm 0.06 (\mathrm{stat}) \pm 0.15 (\mathrm{syst})$ & \\
                            & $\ell^{-}\nu$ & $4.09 \pm 0.05 (\mathrm{stat}) \pm 0.14 (\mathrm{syst})$ & \\\hline
           \multirow{3}*{Z} & $ee$       & $0.960 \pm 0.037 (\mathrm{stat}) \pm 0.059 (\mathrm{syst})$ & \multirow{3}{*}{$0.972 \pm 0.042$} \\
                            & $\mu\mu$   & $0.924 \pm 0.031 (\mathrm{stat}) \pm 0.022 (\mathrm{syst})$ & \\
                            & $\ell\ell$ & $0.931 \pm 0.026 (\mathrm{stat}) \pm 0.023 (\mathrm{syst})$ & \\

       \bottomrule
   \end{tabular}
   \label{tab:xsecResults}
\end{table}

Table~\ref{tab:ratioResults} lists the measured W/Z and W$^{+}$/W$^{-}$ cross section ratios,
which are denoted $R_{\mathrm{W/Z}}$ and $R_{\mathrm{+/-}}$, respectively.
The measured cross section and ratio values are all in agreement with the predictions.

\begin{table}[hbt]
  \footnotesize
   \centering
   \caption{Summary of the cross section ratio measurements and their theoretical predictions}
   \begin{tabular}{cccc}
       \toprule
       \multicolumn{2}{c}{\textbf{Channel}} & \textbf{$\sigma\times\mathcal{B}$ (nb) } & \textbf{NNLO (nb)} \\
       \midrule
           \multirow{3}*{$R_{\mathrm{W/Z}}$} & $e$    & $10.47 \pm 0.42 (\mathrm{stat}) \pm 0.47 (\mathrm{syst})$ & \multirow{3}{*}{$10.74 \pm 0.04$} \\
                                    & $\mu$  & $10.74 \pm 0.37 (\mathrm{stat}) \pm 0.33 (\mathrm{syst})$  & \\
                                    & $\ell$ & $10.64 \pm 0.28 (\mathrm{stat}) \pm 0.29 (\mathrm{syst})$  & \\\hline
           \multirow{3}*{$R_{\mathrm{+/-}}$} & $e$    & $1.434 \pm 0.028 (\mathrm{stat}) \pm 0.082 (\mathrm{syst})$ & \multirow{3}{*}{$1.43 \pm 0.04$} \\
                                    & $\mu$  & $1.433 \pm 0.026 (\mathrm{stat}) \pm 0.054 (\mathrm{syst})$ & \\
                                    & $\ell$ & $1.433 \pm 0.020 (\mathrm{stat}) \pm 0.050 (\mathrm{syst})$ & \\
       \bottomrule
   \end{tabular}
   \label{tab:ratioResults}
\end{table}

\begin{figure}[htb]
   \centering
   \includegraphics*[width=65mm]{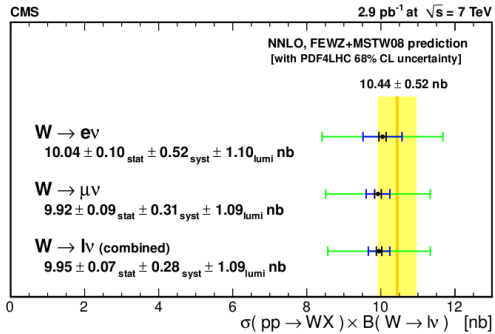}
   \caption{Summary of the W boson production cross section times branching ratio measurements}

   \label{fig:W}
\end{figure}

\begin{figure}[htb]
   \centering
   \includegraphics*[width=65mm]{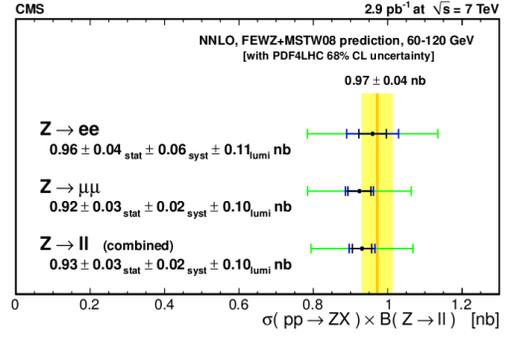}
   \caption{Summary of the Z boson production cross section times branching ratio measurements}

   \label{fig:Z}
\end{figure}

\begin{figure}[htb]
   \centering
   \includegraphics*[width=65mm]{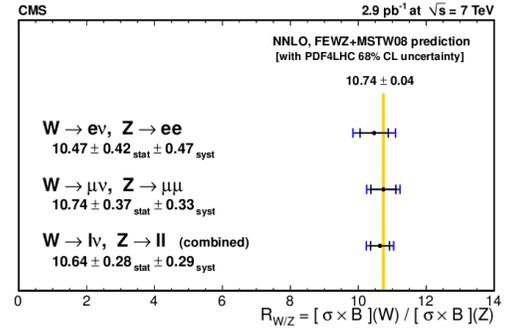}
   \caption{Summary of the $R_{\mathrm{W/Z}}$  cross section ratio measurements}

   \label{fig:R_WZ}
\end{figure}

\begin{figure}[htb]
   \centering
   \includegraphics*[width=65mm]{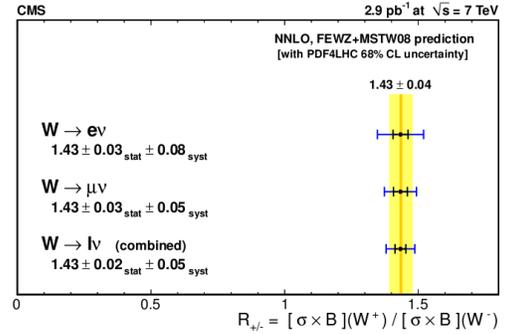}
   \caption{Summary of the $R_{\mathrm{+/-}}$  cross section ratio measurements}

   \label{fig:R_Wpm}
\end{figure}

Summaries of the measurements are given in Figures~\ref{fig:W}, \ref{fig:Z}, \ref{fig:R_WZ}, and \ref{fig:R_Wpm}, illustrating the
consistency of the measurements in the electron and muon channels, as well as the confirmation of theoretical 
predictions computed at the NNLO in QCD with state-of-the-art
PDF sets. For each reported measurement, the statistical error is represented in black and
the total experimental uncertainty, obtained by adding in quadrature the statistical and
systematic uncertainties, in dark blue. For the cross section measurements, the luminosity
uncertainty is added linearly to the experimental uncertainty, and is represented in green.
The dark-yellow vertical line represents the theoretical prediction, and the light-yellow
vertical band is the theoretical uncertainty, interpreted as a 68\% confidence interval.


\section{PROJECTED PRECISIONS FOR A 36 pb$^{-1}$ ANALYSIS}
The overall uncertainty on the W/Z cross section
results will be substantially reduced when measurements are made on the full 36 pb$^{-1}$ dataset,
due to larger event yields, more detailed understanding, and improvements in analysis techniques.
Table~\ref{tab:projections_e} 
compares the statistical and (non-luminosity) systematic errors reported for the 2.9 pb$^{-1}$ analysis to conservative
predictions for the 36 pb$^{-1}$ analysis in the electron and muon channels.  The reduction
suggests using electroweak boson decays as a luminometer will be competitive with VdM scans.
This will further be explored in the following section.

\begin{table}[hbt]
  \footnotesize
   \centering
   \caption{Production Cross Section Uncertainties and their Projections for the Electron and Muon Channels}
   \begin{tabular}{ccccc}
       \toprule
       \textbf{$\Delta\sigma/\sigma$} & \multicolumn{2}{c}{$\mathbf{W \to e\nu}$} & \multicolumn{2}{c}{$\mathbf{Z \to ee}$} \\
           \textbf{(\%)}             & \textbf{2.9 pb$^{-1}$} & \textbf{36 pb$^{-1}$} & \textbf{2.9pb$^{-1}$} & \textbf{36 pb$^{-1}$} \\      
       \midrule
       Stat  & 0.6                     & 0.2                   & 3.8                   & 1.1 \\
       Syst   & 5.1                     & 4.0                   & 6.2                   & 4.2 \\\hline
       Total        & 5.1                     & 4.0                   & 7.3                   & 4.3 \\
       \bottomrule
   \end{tabular}

   \begin{tabular}{ccccc}
       \multirow{2}{*}{\textbf{~~~~}} & \multicolumn{2}{c}{$\mathbf{W \to \mu\nu}$} & \multicolumn{2}{c}{$\mathbf{Z \to \mu\mu}$} \\
                        & \textbf{2.9 pb$^{-1}$} & \textbf{36 pb$^{-1}$} & \textbf{2.9pb$^{-1}$} & \textbf{36 pb$^{-1}$} \\
       \midrule
       Stat & 0.7                     & 0.2                   & 3.1                   & 0.9 \\
       Syst & 3.1                     & 2.2                   & 2.3                   & 2.1 \\\hline
       Total       & 3.4                     & 2.2                   & 3.9                   & 2.3 \\
       \bottomrule
   \end{tabular}

   \label{tab:projections_e}
\end{table}



\section{COMPARISONS OF W/Z vs. VAN DER MEER SCAN CALIBRATION}
Luminosity at CMS is calibrated via Van der Meer scans~\cite{scans},
where horizontal and vertical beam separation scans are performed to measure
the beam sizes.  The beam sizes along with the other known machine parameters 
determine the luminosity. 
The consistency of the results obtained for the cross section measurements
in addition to the copious signal yields and precisely known cross sections, 
suggest the possibility of inverting
the cross section measurements to instead use the signal yield to calibrate the luminosity.
This can be demonstrated with Z bosons.
Table~\ref{tab:projections_lumi} compares the current (with 2.9 pb$^{-1}$) and projected (36 pb$^{-1}$)
systematic uncertainties of a luminosity
calibration using either Z bosons (combined electron and muon channels) or VdM scans.

\begin{table}[hbt]
  \footnotesize
   \centering
   \caption{Luminosity Calibration Systematic Uncertainties and their Projections}
   \begin{tabular}{ccccc}
       \toprule
       \multirow{2}{*}{\textbf{}} & \multicolumn{2}{c}{\textbf{VdM Scan}} & \multicolumn{2}{c}{$\mathbf{Z \to \ell\ell}$} \\
                        & \textbf{2.9 pb$^{-1}$} & \textbf{36 pb$^{-1}$} & \textbf{2.9pb$^{-1}$} & \textbf{36 pb$^{-1}$} \\
       \midrule
       $\Delta\mathcal{L}/\mathcal{L}$ (\%) & 11                     & 4                   & 6                   & 4-5 \\
       \bottomrule
   \end{tabular}
   \label{tab:projections_lumi}
\end{table}

The dominant systematic uncertainty on the VdM scans is the beam current measurement.
On the other hand, the dominant systematic uncertainty on the Z based calibration comes from the PDF~\cite{halyo}.
Table~\ref{tab:projections_lumi} shows that calibrating the luminosity using Z bosons is competitive with the
separation scans.  
It indicates a Z decay based luminosity calibration with a precision of 4-5\% should be possible on a daily basis
if the LHC provides CMS with approximately 30-40 pb$^{-1}$ of collisions per day.
Still, continued improvement of the VdM scans (in particular reducing the uncertainty on the beam current 
measurement) is advocated since scans can be used to constrain the proton PDF.

\section{Z YIELD STABILITY FOR LUMINOSITY CALIBRATION}
\begin{figure}[htb]
   \centering
   \includegraphics*[width=65mm]{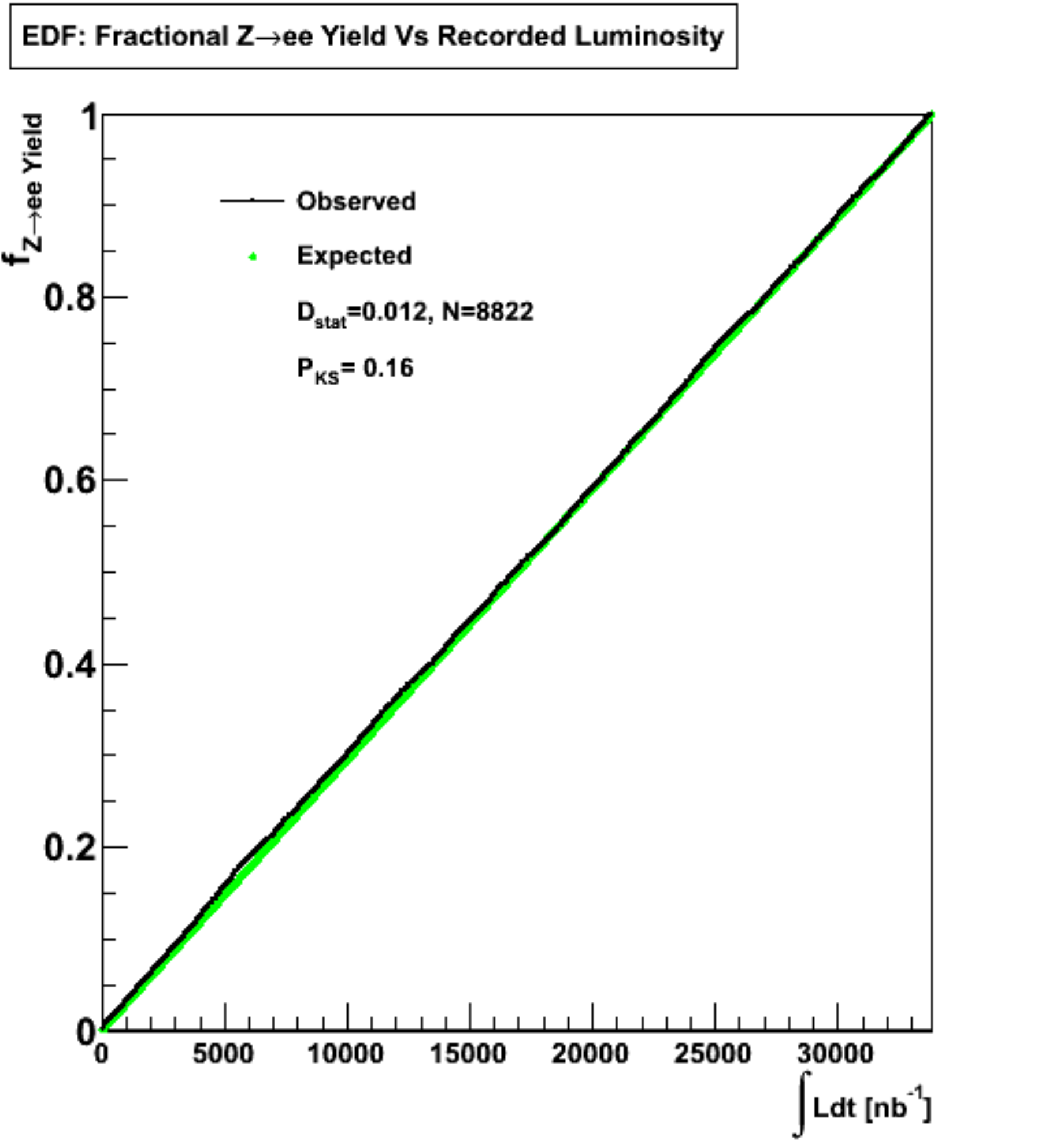}
   \caption{The $Z\to ee$ EDF vs. scan based luminosity}
   \label{fig:ks_test}
\end{figure}

Although VdM scans currently provide the primary method to calibrate the luminosity,
W/Z bosons could be used as a cross check in the coming periods of data taking.
The signal yield must be continuously validated to use these decays for calibration.
Irregularities in the signal yield can be uncovered using a Kolmogrov-Smirnov omnibus test~\cite{KS}.
These tests yield
the significance or probability value of an observed or claimed deviation in a given frequency distribution
from the expected distribution. 
In Kolmogrov-Smirnov tests, the empirical distribution function (EDF) of the signal yield is plotted vs. an orthogonal variable.
The orthogonal variable could be the separation scan based luminosity or the yield of another signal.  The EDF
is just the fractional yield of the signal.  This frequency distribution is then compared to the expected
distribution (e.g. the yield increasing linearly with luminosity).  The maximal vertical difference
between the observed and expected distributions determines a probability .  Such a test
is shown in Figure~\ref{fig:ks_test} where the EDF of the $Z\to ee$ yield observed at CMS
is plotted vs. the scan based luminosity for 36 pb$^{-1}$.  The data is shown in black, while the expected
distribution is in green.  The maximal difference between the 2 distributions is labeled on the
plot as $D_{stat}$ and the corresponding probability value is labelled as $P_{KS}$. 
The observed distribution agrees well with the expectation and the high probability value indicates a stable signal yield,
consistent with the hypothesis that the yield increases linearly with luminosity.

\begin{figure}[htb]
   \centering
   \includegraphics*[width=65mm]{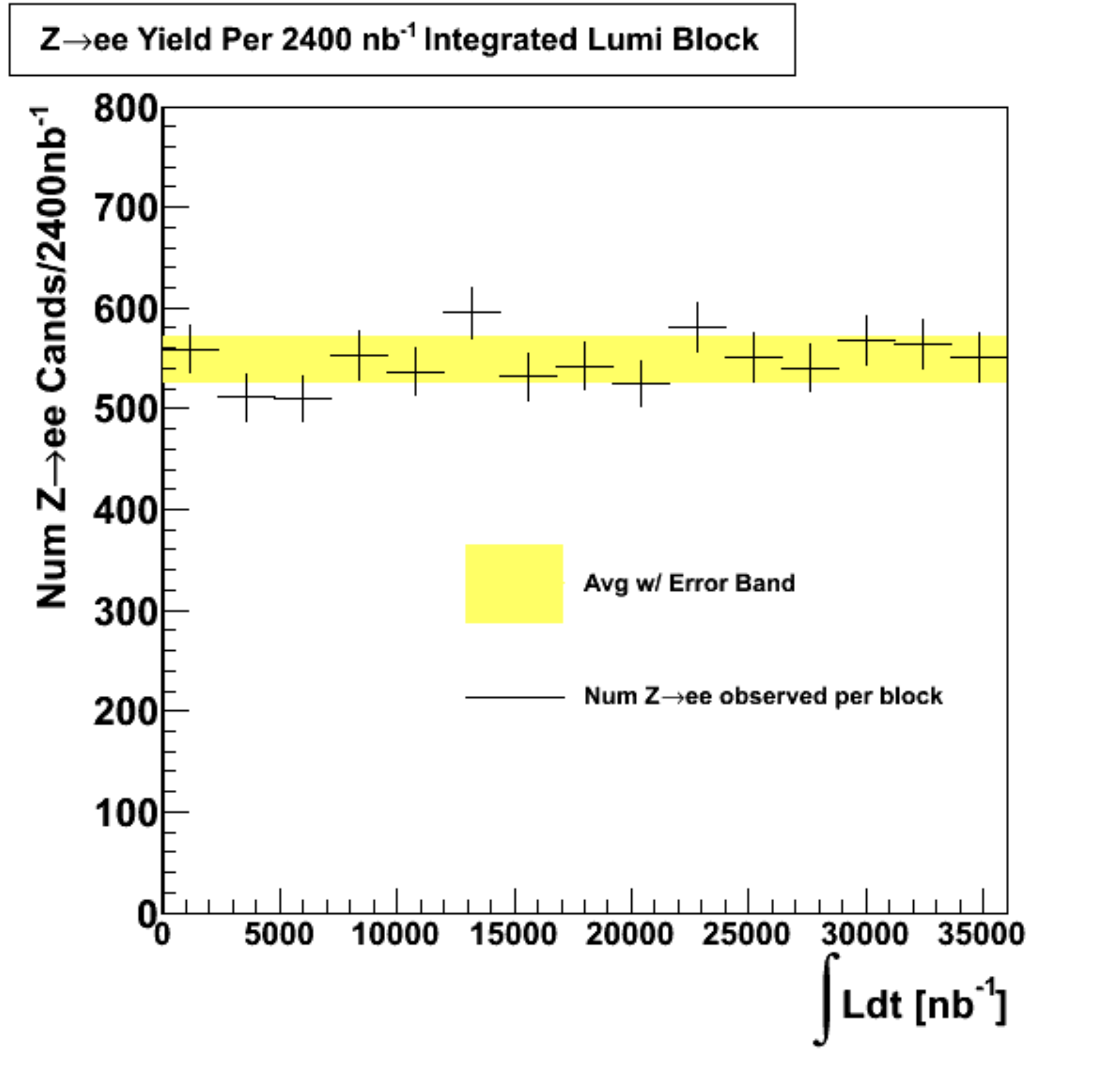}
   \caption{The $Z\to ee$ yield vs. blocks of fixed integrated luminosity}
   \label{fig:stab_eye}
\end{figure}

Other useful checks include plotting the signal yield vs. blocks of fixed integrated luminosity.  One
can then verify that this distribution is flat, having its points agreeing within
errors.  Such a plot is shown in Figure~\ref{fig:stab_eye}, where the $Z\to ee$ yield observed at CMS
is plotted in 2.4 pb$^{-1}$ luminosity blocks with about 4\% relative statistical error per point.
The data is shown in black while the corresponding statistical error
band is plotted in yellow.  The distribution is flat and the yield was stable during data taking.

\section{CONCLUSIONS}
CMS has performed measurements of inclusive W and Z production cross sections in pp collisions at $\sqrt{s} = 7$ TeV
using $2.9\pm0.3$ pb$^{-1}$ of data recorded by the CMS detector at the LHC~\cite{theArticle}.  The W/Z and W$^{+}$/W$^{-}$ production cross 
section ratios were also reported.  Measurements were performed for both electron and muon decay channels, and were
then combined.  The measurements are internally consistent and agree well with the theoretical predictions.
Conservative systematic uncertainty projections for an upcoming 36 pb$^{-1}$ measurement~\cite{EWK10005} were provided~\footnote{
Errors reported in~\cite{EWK10005} are even less: 2.1\% for Z production.}and the feasibility of 
using W/Z boson decays alongside the VdM scans to calibrate the luminosity has been examined.
Such a calibration is possible for 2011 high luminosity data taking,
and usage of W/Z bosons as a standard candle for luminosity could happen as early as this year.



\end{document}